\documentclass[structabstract]{aa}  
\usepackage{graphicx}
\usepackage{txfonts}
\usepackage{natbib}
\begin{document}

\title{High-mass star formation at sub-50\,AU scales \thanks{The data
    are available in electronic form at the CDS via anonymous ftp to
    cdsarc.u-strasbg.fr (130.79.128.5) or via
    http://cdsweb.u-strasbg.fr/cgi-bin/qcat?J/A+A/}.}


   \author{H.~Beuther
          \inst{1}
          \and
          A.~Ahmadi
          \inst{1}
           \and
          J.C.~Mottram
          \inst{1}
          \and
          H.~Linz
          \inst{1}
          \and
          L.T.~Maud
          \inst{2,7}
          \and
          Th.~Henning
          \inst{1}
          \and
          R.~Kuiper
          \inst{4, 1}
          \and
          A.J.~Walsh
          \inst{3}
          \and
          K.G.~Johnston
          \inst{5}
          \and
          S.N.~Longmore
          \inst{6}
}
   \institute{$^1$ Max Planck Institute for Astronomy, K\"onigstuhl 17,
              69117 Heidelberg, Germany, \email{beuther@mpia.de}\\
              $^2$ Leiden Observatory, Leiden University, PO Box 9513, NL-2300 RA Leiden, the Netherlands\\
              $^3$ {Research Centre for Astronomy, Astrophysics, and Astrophotonics, Macquarie University, NSW 2109, Australia}\\
              $^4$ Institute of Astronomy and Astrophysics, University of T\"ubingen, Auf der Morgenstelle 10, D-72076 T\"ubingen, Germany\\
              $^5$ School of Physics and Astronomy, University of Leeds, Leeds, LS2 9JT, UK\\
              $^6$ Astrophysics Research Institute, Liverpool John Moores University, 146 Brownlow Hill, Liverpool L3 5RF, UK\\
              $^7$ European Southern Observatory, Garching bei Munchen, Germany
}

   \date{Version of \today}

\abstract
{The hierarchical process of star formation has so far mostly been 
  studied on scales from thousands of au to parsecs, but the smaller
  sub-1000\,au scales of high-mass star formation are still largely
  unexplored in the submm regime.}
{We aim to resolve the dust and gas emission at the highest spatial
  resolution to study the physical properties of the densest
  structures during high-mass star formation.}
{We observed the high-mass hot core region G351.77-0.54 with the Atacama
  Large Millimeter Array with baselines extending out to more than
  16\,km. This allowed us to dissect the region at sub-50\,au spatial
  scales.}
{At a spatial resolution of 18/40\,au (depending on the distance), we
  identify twelve sub-structures within the inner few thousand au of
  the region. The brightness temperatures are high, reaching values
  greater 1000\,K, signposting high optical depth toward the peak
  positions. Core separations vary between sub-100\,au to several 100
  and 1000\,au. The core separations and approximate masses are
  largely consistent with thermal Jeans fragmentation of a dense gas
  core. Due to the high continuum optical depth, most spectral lines
  are seen in absorption. However, a few exceptional emission lines
  are found that most likely stem from transitions with excitation
  conditions above 1000\,K. Toward the main continuum source, these
  emission lines exhibit a velocity gradient across scales of
  100-200\,au aligned with the molecular outflow and perpendicular to
  the previously inferred disk orientation. While we cannot exclude
  that these observational features stem from an inner hot accretion
  disk, the alignment with the outflow rather suggests that it stems
  from the inner jet and outflow region. The highest-velocity features are
  found toward the peak position, and no Hubble-like velocity
  structure can be identified. Therefore, these data are consistent
  with steady-state turbulent entrainment of the hot molecular gas via
  Kelvin-Helmholtz instabilities at the interface between the jet and
  the outflow.}
{Resolving this high-mass star-forming region at sub-50\,au scales
  indicates that the hierarchical fragmentation process in the
  framework of thermal Jeans fragmentation can continue down to the
  smallest accessible spatial scales. Velocity gradients on these
  small scales have to be treated cautiously and do not necessarily
  stem from disks, but may be better explained with outflow
    emission. Studying these small scales is very powerful, but
    covering all spatial scales and deriving a global picture from
    large to small scales are the next steps to investigate.}

\titlerunning{High-mass star formation at sub-50\,au scales}

\maketitle

\section{Introduction}
\label{intro}

Star formation is an intrinsically hierarchical process where
fragmentation occurs on almost all scales, from the large-scale
molecular clouds with sizes of several ten to hundred pc, down to
close binary formation at the center of dense cores and disk
fragmentation that can lead to planets. In that framework, the
smallest scale fragmentation of the cores and potentially also the
disks are poorly understood, in particular for regions forming
high-mass stars ($>8$\,M$_{\odot}$) because they are typically located
at distances of several kpc. Until recently, the state-of-the-art
observations had as high resolution as $\sim$0.2$''$, allowing for the
investigation of the fragmentation processes of the larger-scale
molecular gas clumps into their cluster-forming cores (e.g.,
\citealt{beuther2007d,bontemps2010,palau2013,palau2014,beuther2013b,csengeri2017,cesaroni2017,beuther2018b}). To understand the formation of close binaries
or systems like the Trapezium in Orion (e.g.,
\citealt{preibisch1999}), higher spatial resolution is
needed. Furthermore, many other physical processes take place on small
spatial scales. For example, disk-like structures as well as the inner
jet and outflow regions are found on scales on the order of 10 to a few
100\,au (e.g., \citealt{tan2014,frank2014,beltran2016}), again
requiring the highest spatial resolution possible. Additionally important are global infall and streamer-like motions through which the material can get channeled in the innermost regions (e.g., \citealt{maud2017,goddi2018}).

\begin{figure*}[htb]
  \includegraphics[width=0.49\textwidth]{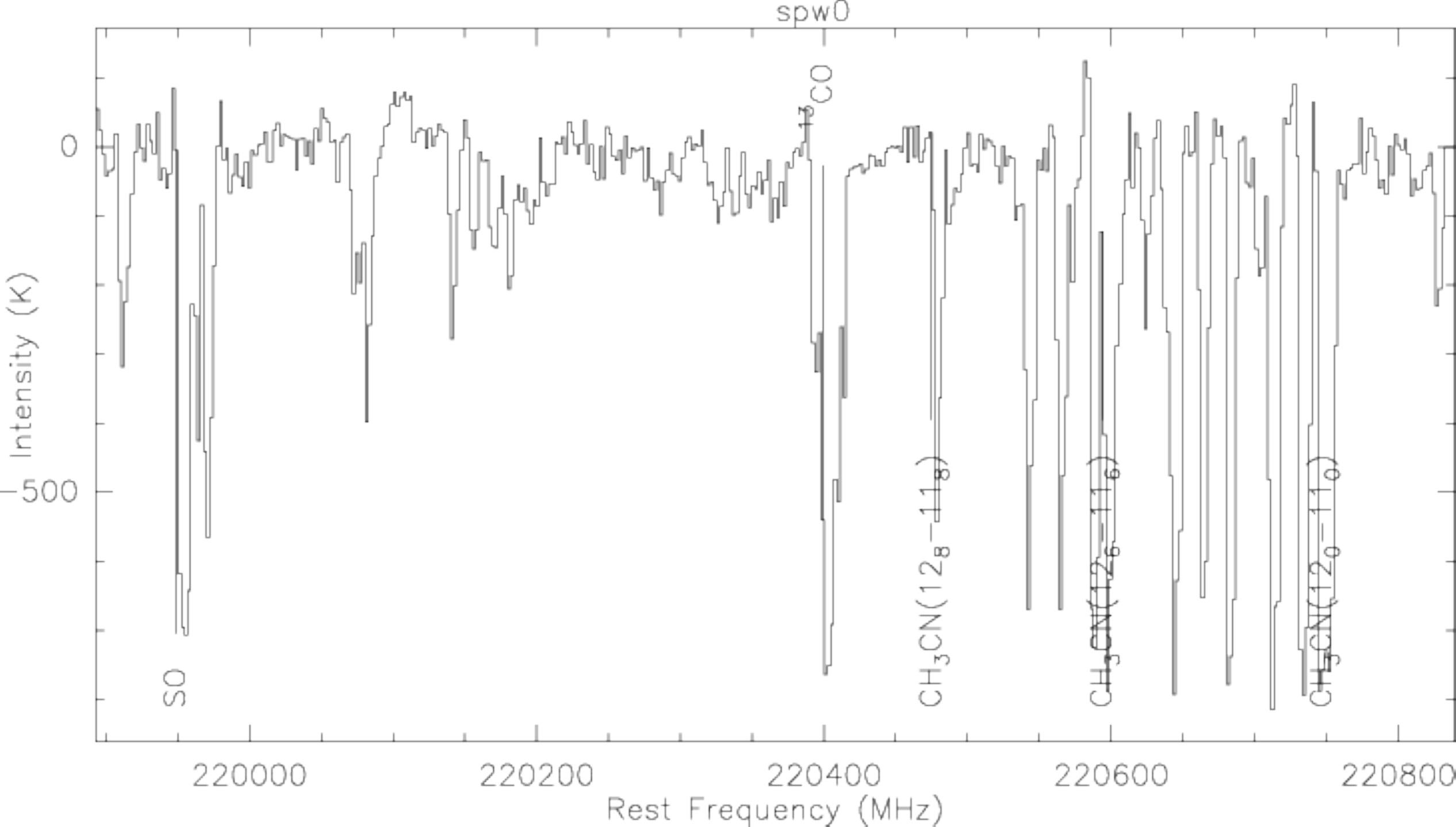}
  \includegraphics[width=0.49\textwidth]{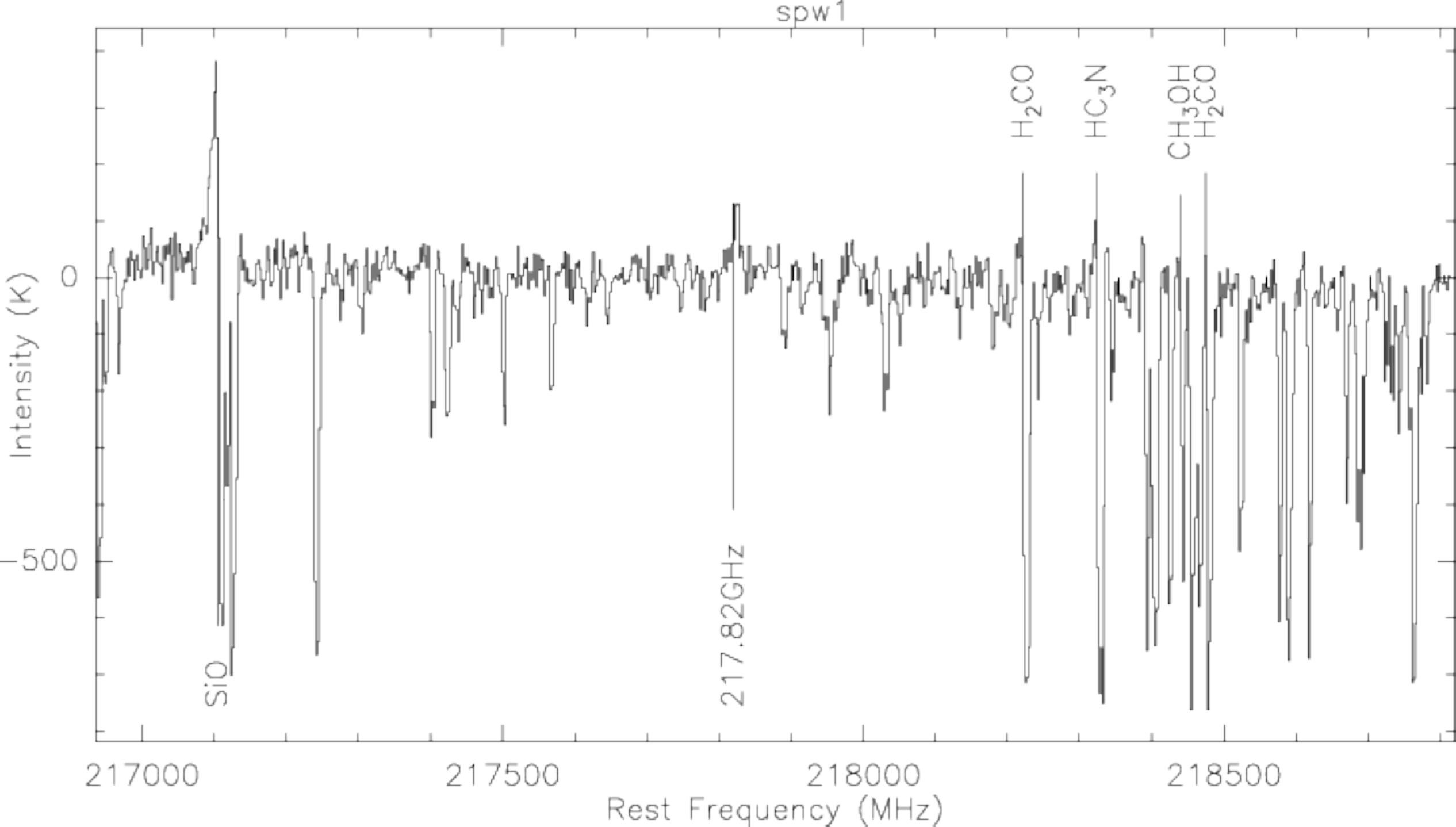}\\
  \includegraphics[width=0.49\textwidth]{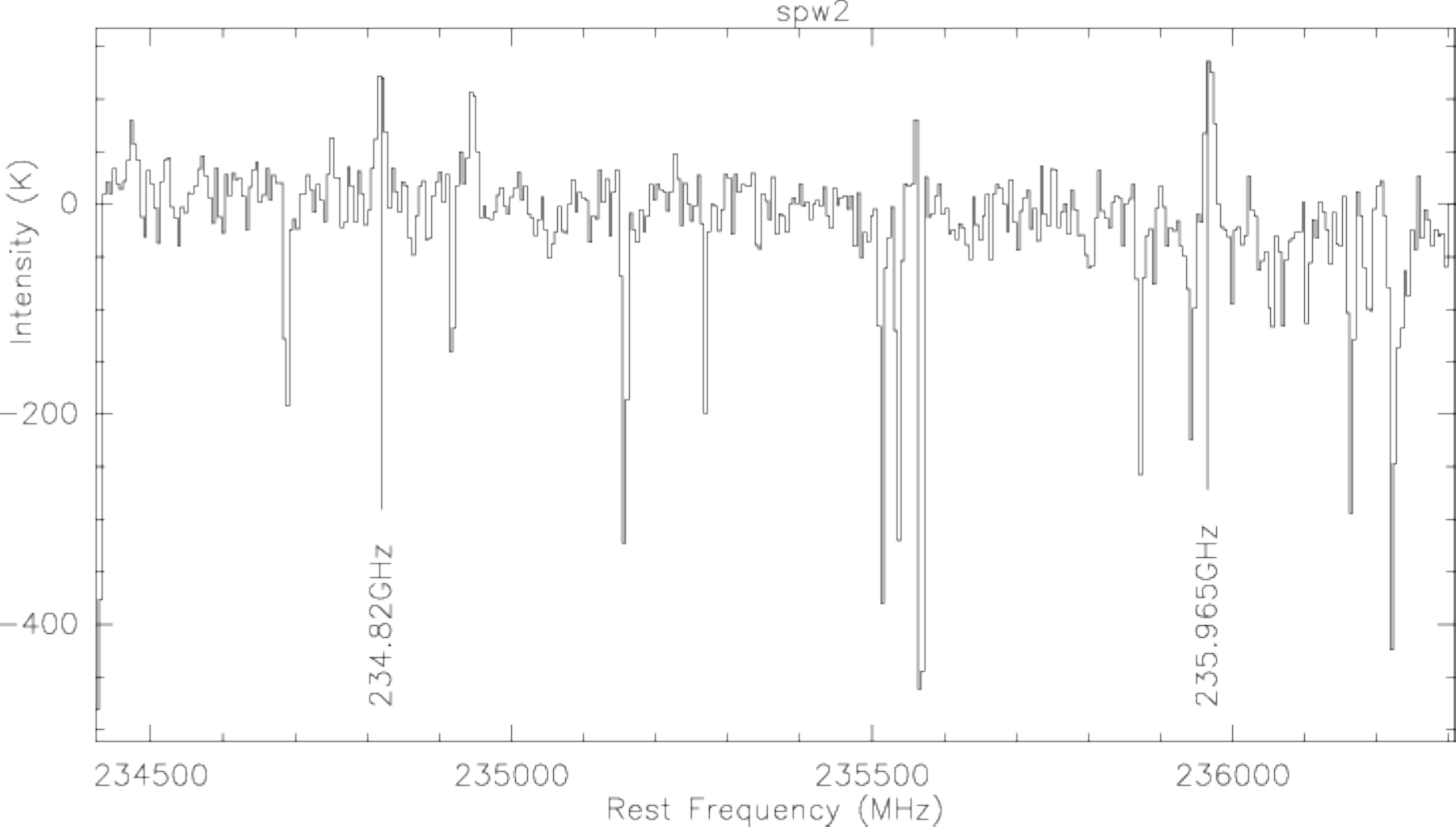}
  \includegraphics[width=0.49\textwidth]{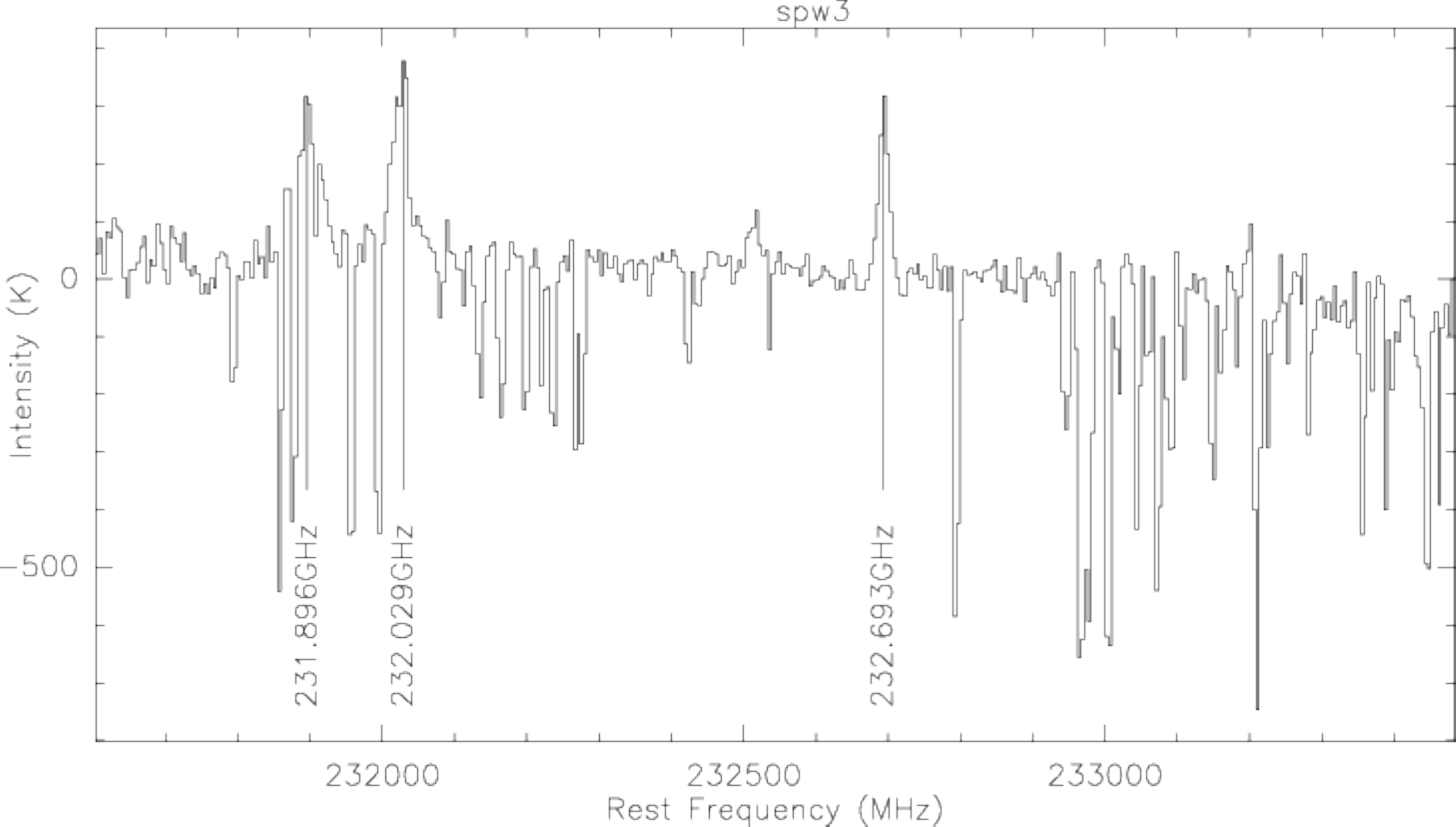}
  \caption{Spectra of the four spectral windows extracted toward the
    main mm continuum peak position mm1. Clearly most lines are seen in
    absorption against the continuum, only a few in emission. Selected
    spectral lines are labeled.}
\label{spectra} 
\end{figure*} 
                                                
Millimeter and centimeter wavelengths are the ideal spectral regime to
study these processes because the young disks and jets are still
deeply embedded within their natal cores. Sub-$0.1''$ spatial
resolution can be achieved with the Very Large Array at cm wavelengths
(e.g., \citealt{moscadelli2014,beuther2017c}), and now also with the
Atacama Large Millimeter Array (ALMA) at (sub)mm wavelengths (e.g.,
\citealt{brogan2015}). 

Our target region is the well-known high-mass star-forming region
G351.77-0.54 (a.k.a.~IRAS\,17233-3606). The distance to the target is
debated, and while early studies put it at 2.2\,kpc
\citep{norris1993}, more recent investigations favor smaller distances
around 1\,kpc \citep{leurini2011}. Since the accurate distance is not
clear yet, in the following, we will always give the parameters for
both distances. The luminosity and mass of the entire region are
estimated to be $1.7\times 10^4/10^5$\,L$_{\odot}$ and
660/3170\,M$_{\odot}$ at 1/2.2\,kpc, respectively
\citep{norris1993,leurini2011,leurini2011b}. These numbers clearly
show that we are dealing with a massive star-forming region on its way
to form high-mass stars. The region exhibits linear CH$_3$OH class II
maser features as well as strong outflow emission (e.g.,
\citealt{norris1993,walsh1998,leurini2009,leurini2013,klaassen2015}).
Furthermore, this region is a very bright (sub)mm line emission
source. Centimeter-wavelength interferometric studies revealed strong
NH$_3$(4,4)/(5,5) emission at temperatures above 200\,K and resolving
overall rotational motion on a few 1000\,au scales
\citep{beuther2009c}, approximately perpendicular to the larger-scale
outflow. Following \citet{leurini2011}, we use as $\varv_{\rm lsr}$ a
value of $\sim -3.6$\,km\,s$^{-1}$. Since \citet{zapata2008} found six
cm continuum sources within that core, it was expected that this
region would show fragmentation on smaller spatial scales in the
(sub)mm regime. Furthermore, multiple outflows have been identified
(e.g.,
\citealt{leurini2009,leurini2011,klaassen2015,beuther2017b}). In a
recent ALMA study of this region at 690\,GHz we resolved the region at
0.06$''$ resolution \citep{beuther2017b}. This study revealed at least
four sub-cores where the central source \#1 provide strong evidence
for rotational motion on scales of a few hundred au,
perpendicular to an outflow seen in CO(6--5). But ALMA can go to even
higher spatial resolution, and here we present a study of the same
region at $\sim$1.3\,mm wavelengths and 21\,mas$\times$15\,mas
resolution, corresponding to $\sim$18/40\,au at a distance of
1/2.2\,kpc.

\section{Observations} 
\label{obs}

The target high-mass star-forming region G351.77-0.54 was observed
with ALMA as a cycle 3 project (id 2015.1.00496) in four separate
observing blocks between Nov.~5 and Dec.~2, 2015. The duration of the
four scheduling blocks varied between 1 and 1.6 hours with on-source
target observing times between 24 and 45 minutes. Between 33 and 43
antennas were in the array and baselines between 17\,m and more than
16\,km were covered. The phase center of our observations was
R.A.~(J2000.0) 17:26:42.568 and Dec.~(J2000.0) -36:09:17.60. To
calibrate the phases on the long baselines, short cycle times were
used between the target and the phase and amplitude calibrator
J1733-3722 with a typical cycle times consisting of $\sim$55\,sec on
source and $\sim$18\,sec on the calibration quasar. Bandpass and flux
calibration were conducted with the sources J1517-2422 and J1617-5848,
respectively. The precipitable water vapor (PWV) varied between 0.42
and 0.93\,mm. The 1.3\,mm band 6 receivers were tuned to the LO
frequency of 226.378\,GHz with two spectral windows (spws) in the
upper and lower sidebands, respectively. The frequency coverages of
the four spectral windows is depicted in Fig.~\ref{spectra}. The
spectral resolution of the spws was 244\,kHz, 488\,kHz, and for the
wider spectral windows 3906\,kHz, corresponding to velocity
resolutions of $\sim$0.33, $\sim$0.67 and 5.3\,km\,s$^{-1}$,
respectively. The primary beam of ALMA at these frequencies is
$\sim$27$''$.

\begin{figure*}[htb]
  \includegraphics[width=1.0\textwidth]{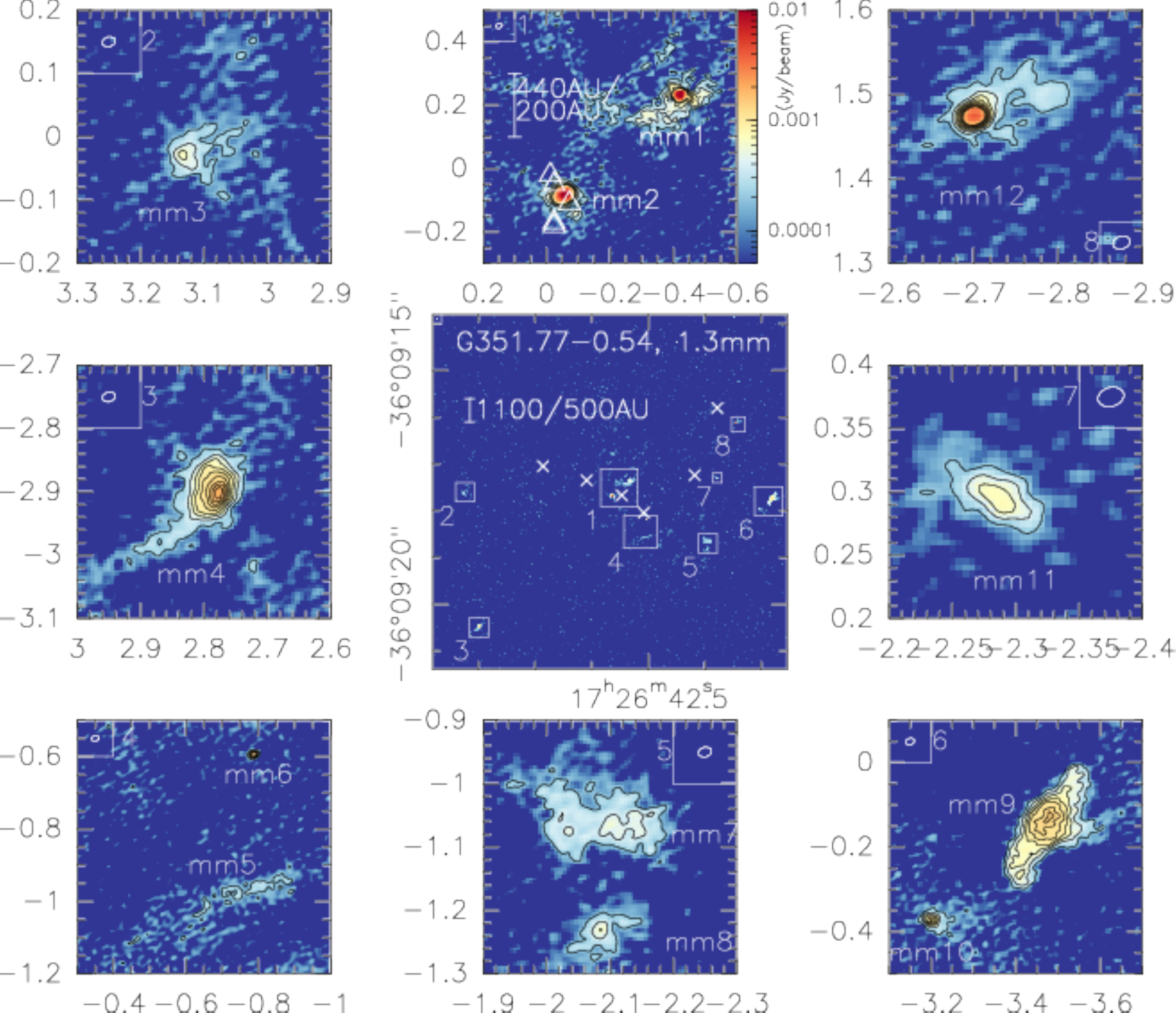}
  \caption{ALMA 1.3\,mm continuum data toward the high-mass
    star-forming region G351.77-0.54 at an angular resolution of
    $0.021''\times 0.015''$. The central panel shows the overall
    structure (the white crosses mark the cm continuum source
    positions from \citealt{zapata2008}), and the surrounding
    panels show selected zooms into the areas marked in the central
    panel. We note the different cutout sizes of the individual zoom
    windows. The contour levels always start at a $4\sigma$ level of
    0.2\,mJy\,beam$^{-1}$ and continue in $4\sigma$ steps. Scale bars
    assuming 1.0/2.2\,kpc distance are shown in the middle and
    middle-top panels. The middle-top panel also shows the CH$_3$OH
    maser positions from \citet{walsh1998} and an intensity scale. The
    synthesized beam is presented in each panel.}
\label{cont} 
\end{figure*} 

The calibration of the data was done with the ALMA pipeline in
\textsc{CASA} \citep{mcmullin2007} version 4.5.0 following the scripts
provided by the observatory. In a following step, we applied
phase-only self-calibration to the data going down to time-steps of
30\,sec. Images prior and after self-calibration were carefully
compared, and while all structures are well recovered, the noise is
decreased after self-calibration. No amplitude self-calibration was
applied.

The following imaging was also conducted within
\textsc{CASA}. Although some baselines between 17 and 55\,m are
observed, their coverage is sparsely sampled, and so do not allow us
to reasonably image the larger-scale emission. Therefore, in the
imaging process we only used data from baselines longer than 55\,m,
which allows us to concentrate on the most compact structures. As
visible in Fig.~\ref{spectra}, this hot core hosts many spectral lines
and defining line-free parts of the spectrum is not easy. Therefore,
we compared different approaches to create a continuum image of the
region. On the one hand, we produced a continuum image from only a
small part of the spectrum that appears line-free, and on the other
hand, we produced a continuum image from the whole bandpass with all
spectral features included. Structurally and with respect to the
intensities, there was barely any difference between the two images,
only the noise of the full-bandpass image was lower. Therefore, in the
following we use the continuum image created from the whole
bandpass. For the continuum images we applied a robust weighting of
-2, corresponding to uniform weighting. This results in a synthesized
beam of 21\,mas\,$\times$\,15\,mas with a position angle of
-74$^{\circ}$. A few apparently line-free parts of the spectra were
used for the continuum subtraction. However, in such a line-rich
source, this is not a trivial process and some weak bandpass slope
remains (Fig.~\ref{spectra}). However, that slope is so weak that is
does not affect any of the analyses. The line data were then imaged
with a robust weighting of 0, resulting in a slightly larger
synthesized beam of $\sim$25\,mas\,$\times$\,20\,mas with a position
angle of -86$^{\circ}$. The continuum 1$\sigma$ rms is
$\sim$50\,$\mu$Jy\,beam$^{-1}$, and the typical rms for the spectral
lines in a line-free channel of 1\,km\,s$^{-1}$ width is
$\sim$0.9\,mJy\,beam$^{-1}$.

\section{Results}

As outlined in the previous Section \ref{cont} and shown in
Fig.~\ref{spectra}, extended emission is not well imaged in this
long-baseline-only dataset, and most lines are mainly seen in
absorption against the continuum sources. However, there are a few
emission lines (Fig.~\ref{spectra}) showing structure on small spatial
scales below 100\,au. Therefore, in the following we concentrate on
the small-scale continuum emission as well as on the few transitions
seen in emission.

\subsection{Continuum emission and core fragmentation}
\label{sec_cont}

Figure \ref{cont} presents a compilation of the 1.3\,mm continuum data
of the region. While the central panel shows the inner $\sim 7''\times
7''$, the surrounding 8 panels present zoom-ins for the
sub-structure. With an angular resolution of $0.021''\times 0.015''$
(mean of $0.018''$), we can resolve linear structures of roughly
18/40\,au at 1/2.2\,kpc distance, among the highest spatial resolution
observations achieved for high-mass star-forming regions today.

While at first sight, this image appears largely empty with only
a few compact sub-structures distributed throughout, one should keep
in mind that this image is strongly affected by missing flux because
we only image data with baselines larger than 55\,m (and extending to
more than 16\,km). Hence, all intermediate and large-scale structures
are filtered out and we concentrate on the small sub-cores
remaining. To get a feeling for the fraction of missing flux, we
extract the 870\,$\mu$m flux (central frequency 345\,GHz) of
G351.77-0.54 from the ATLASGAL survey \citep{schuller2009}. Scaling
the 870\,$\mu$m peak flux density of 52.1\,Jy\,beam$^{-1}$ (with a
beam size of $19.2''$) with a flux dependency of $S\propto \nu^4$ to
our central frequency of 226.6\,GHz, approximately 9.7\,Jy should be
emitted in our field. However, as shown below, the sum of the fluxes
of all identified sub-cores accumulates to only $\sim$208\,mJy. Taking
these numbers, we recover only about 2.1\% of the total continuum
emission in the field.

\begin{table*}[htb]
\caption{Millimeter continuum emission parameters}
\label{mm_continuum}
\begin{tabular}{l|rr|rrrr|rrr}
\hline 
\hline
name & R.A. & Dec. & $S_{\rm peak}$ & $S_{\rm int}$ & $T_{\rm b}$ & mean\,$T_{\rm b}$& M@1.0kpc$^a$ & M@2.2kpc$^a$ & N$^b$ \\
     & (J2000.0) & (J2000.0) & (mJy/beam) & (mJy) & (K) & (K) & (M$_{\odot}$)& (M$_{\odot}$) & $(\times 10^{25}$cm$^{-2}$) \\
\hline
mm1  & 17:26:42.533 & -36:09:17.37 & 15.0 & 47.4 & 1144 & 53 & 0.41 & 2.0 & 3.3 \\
mm2  & 17:26:42.563 & -36:09:17.68 & 13.5 & 42.3 & 1027 & 93 & 0.20 & 1.0 & 3.0 \\
mm3  & 17:26:42.827 & -36:09:17.63 & 0.7  & 4.5  & 56   & 24 & 0.10 & 0.5 & 1.6  \\
mm4  & 17:26:42.797 & -36:09:20.50 & 1.9  & 21.0 & 152  & 44 & 0.22 & 1.1 & 4.5 \\
mm5  & 17:26:42.506 & -36:09:18.58 & 0.3  & 3.6  & 29   & 20 & 0.10 & 0.5 & 0.7 \\
mm6  & 17:26:42.503 & -36:09:18.19 & 1.4  & 1.4  & 111  & 50 & 0.01 & 0.1 & 3.2 \\
mm7  & 17:26:42.395 & -36:09:18.67 & 0.6  & 13.8 & 49   & 26 & 0.27 & 1.3 & 1.4 \\
mm8  & 17:26:42.396 & -36:09:18.83 & 0.6  & 3.0  & 52   & 24 & 0.06 & 0.3 & 1.4 \\
mm9  & 17:26:42.280 & -36:09:17.73 & 1.4  & 55.3 & 115  & 50 & 0.51 & 2.5 & 3.4 \\
mm10  & 17:26:42.304 & -36:09:17.97 & 1.3  & 3.7  & 101  & 32 & 0.06 & 0.3 & 2.9 \\
mm11 & 17:26:42.379 & -36:09:17.31 & 0.8  & 2.9  & 62   & 30 & 0.05 & 0.2 & 1.8 \\
mm12 & 17:26:42.345 & -36:09:16.13 & 4.0  & 10.5 & 310  & 38 & 0.13 & 0.6 & 9.4 \\
\hline                                                           
\hline
\end{tabular}
~\\
Notes:\\
$^a$ Masses are calculated using their mean brightness temperature $T_{\rm b}$.\\
$^b$ Column densities are calculated using the peak brightness temperature $T_{\rm b}$.
\end{table*}

Nevertheless, this spatial filtering also has strong advantages
because we can isolate the compact structures within this region.
With the sparse spatial sampling on short and intermediate baselines,
and only a limited amount of emission features in the map, an
automized source identification is not required. Therefore, we
identify the source structures by eye following the 5$\sigma$ contours
in Fig.~\ref{cont}. This way, we identify 12 separate mm sources
within a radius of roughly $3.5''$ or 3500/7700\,au at 1.0/2.2\,kpc
distance, respectively. Projected separations range from more than
1000\,au to several 100\,au. Some sources, for example, mm1 and mm2, show
even smaller secondary peaks on scales below 100\,au. The largest
extended structures we can trace with these observations are found
toward the sub-sources mm1, mm4 and mm8, and depending on the distance
to the region, these sources have sizes between roughly 200 and
600\,au. Taking mm9 as an example, and assuming the closer distance of
1\,kpc, our spatial resolution corresponds to $\sim$18\,au, and the
extent of the source is roughly 300\,au. This outlines the approximate
spatial dynamic range covered by these data. None of the mm continuum
sources are directly spatially associated with the cm continuum
sources reported in \citet{zapata2008} and marked in
Fig.~\ref{cont}. In comparison to the 438\,$\mu$m submm continuum
sources presented in \citet{beuther2017b}, the here identified sources
mm1 and mm2 correspond to their sources \#1 and \#2. The sources \#3
and \#4 in \citet{beuther2017b} correspond to the mm sources mm12 and
mm3, respectively. While the compact sources most likely host
protostars, some of the more diffuse structures like mm5 or mm7 may
not be of protostellar nature but could rather be remnants of the
larger-scale envelope where only some imaged structure is left after
the interferometric spatial filtering (see also Section
\ref{fragmentation}).

At these spatial scales, optical depth effects can also become
important, and one way to assess this is by converting the peak flux
densities into peak brightness temperatures. Table \ref{mm_continuum}
presents for all 11 identified sub-structures the peak flux densities
$S_{\rm peak}$, the integrated fluxes $S_{\rm int}$ (integrated within
the 4$\sigma$ contours), the corresponding peak brightness temperature
$T_{\rm b}$ as well as the mean brightness temperatures derived again
within the 4$\sigma$ contours of 16\,K. As one sees, the peak
brightness temperatures toward the strongest positions mm1 and mm2
exceed 1000\,K, and toward the other peak positions are between 29
and 310\,K. From the CH$_3$CN temperature analysis for this region at
690\,GHz presented in \citet{beuther2017b}, we know that the gas
temperatures around mm1 are also around 1000\,K. This clearly shows
that the dust emission in this region is optically thick and traces an
inner core surface at the derived brightness temperature. Such high
dust optical depth can also be inferred from the fact that almost all
lines are seen in absorption against the mm1 continuum peak
(Fig.~\ref{spectra}). While the peak brightness temperatures in the
other sub-structures are lower, they are still high compared to
studies at slightly lower angular resolution (e.g., the NOEMA CORE
survey at $\sim 0.4''$, \citealt{beuther2018b}).

What gas temperature would one expect for the outer sub-structures
assuming the heating is dominated by the innermost core mm1? In this
case we can do a simple estimate assuming a typical temperature
distribution in dust and gas cores of $T\propto r^{-0.4}$ (e.g.,
\citealt{vandertak2000}). Using 1000\,K as the temperature at our
resolution element (mean value $0.018''$), at the approximate spatial
separation of the outer core structures of $3.5''$ one would get a
temperature of $\sim$121\,K. Taking into account that (a) the medium
is not homogeneous, (b) there are probably several heating sources,
and that (c) the 1000\,K at the resolution element is an lower limit
because it is an average over the beam size, the measured peak
brightness temperatures do not deviate too much from the temperatures
one could expect on these scales in such a region if they were heated
by an internal source like mm1. It is likely that some of the lower
intensity structures already have moderate optical depth at such high
spatial resolution as well.

\begin{figure}[htb]
  \includegraphics[width=0.49\textwidth]{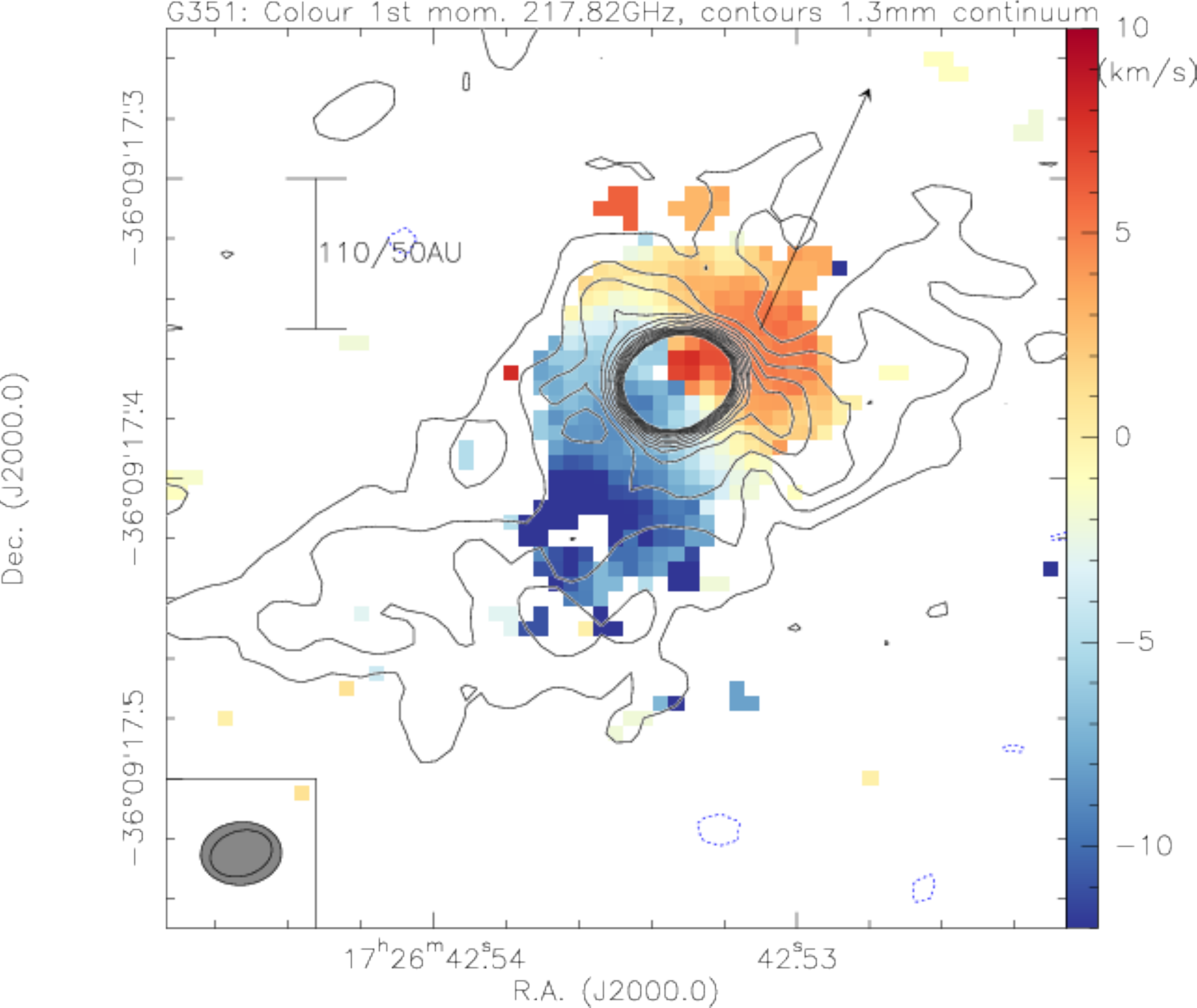}
  \caption{Color-scale shows for mm1 the first moment map
    (intensity-weighted peak velocities) from the line at
    217.82\,GHz. The contours are the 1.3\,mm continuum data in
    $4\sigma$ level of 0.2\,mJy\,beam$^{-1}$ up to a
    3\,mJy\,beam$^{-1}$ level (to avoid too many contours in the
    center). A scale-bar and the resolution elements of the line
    (gray) and continuum are shown as well. The arrow outlines the
    approximate direction of the CO(6--5) red outflow lobe shown in
    Fig.~\ref{mom1_co}.}
\label{mom1} 
\end{figure} 

Although we argue that the dust emission has increased optical depth
at several positions, we nevertheless would like to get rough
estimates of the masses and column densities toward the individual
sub-structures. Therefore, we use the typical approach to estimate
masses from mm continuum emission assuming optically thin emission
(e.g., \citealt{hildebrand1983}), but at the same time stress that the
derived masses and column densities should be considered as lower
limits. Following the equations outlined in \citet{schuller2009} for
dust emission, we are using for the 1.3\,mm dust opacity $\kappa$ a
value of 1.11\,cm$^2$\,g$^{-1}$ from \citet{ossenkopf1994} at high
densities of $10^8$\,cm$^{-3}$. Furthermore, a gas-to-dust mass ratio
of 150 is assumed \citep{draine2011}. Regarding the temperatures,
since we are apparently dealing with optically thick emission, using
the derived brightness temperature as an approximate proxy to estimate
the actual temperatures seems a reasonable approach. For the column
density estimates, we use 1000\,K for the two very bright sources mm1
and mm2, and lower temperatures of 100\,K for the other sources
(depending on the dust optical depth peak brightness temperatures for
some of the weaker sources may be lower limits for the actual gas
temperature). Since the substructures are extended and cover a range
of brightness temperatures, for the mass estimates, we derive for each
core a mean brightness temperature within the $4\sigma$ contour (Table
\ref{mm_continuum}).

\begin{figure*}[htb]
  \includegraphics[width=1.0\textwidth]{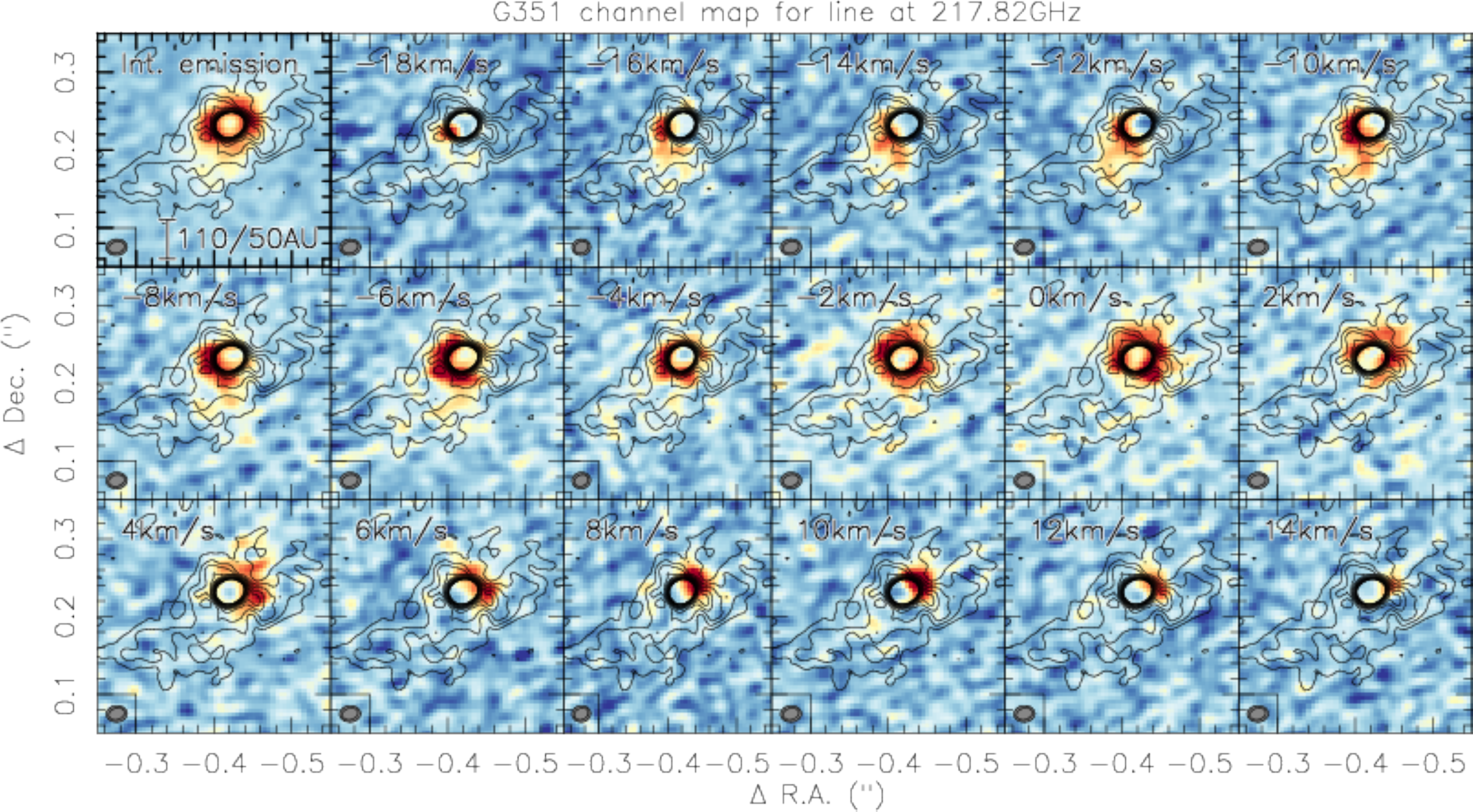}
  \caption{Channel map of the 217.82\,GHz line for mm1. The top-left
    panel presents the integrated emission between -18 and
    +14\,km\,s$^{-1}$. The remaining panels show the channels with
    2\,km\,s$^{-1}$ width at the velocities marked at the top-left of
    each panel. The contours are the 1.3\,mm continuum data in
    $4\sigma$ level of 0.2\,mJy\,beam$^{-1}$ up to a
    3\,mJy\,beam$^{-1}$ level (to avoid too many contours in the
    center). A scale-bar is shown in the top-left panel, and the
    resolution elements of the line (gray) and continuum are presented
    in each panel. The $\varv_{\rm lsr}$ is at approximately
    $-3.6$\,km\,s$^{-1}$.}
\label{channel} 
\end{figure*} 

Table \ref{mm_continuum} presents the derived masses and column
densities, and we stress that these values have to be considered as
lower limits because of the high optical depth as well as the large
amount of missing flux. Hence, although individual core masses between
a fraction of a solar mass and 2\,M$_{\odot}$ (depending on distance)
do not seem very high, since they are lower limits and measured over
very small size scales, they already indicate very high
densities. For example, the almost spherical source mm2 has
at 1\,kpc distance a mass lower limit of 0.2\,M$_{\odot}$ and a
diameter of $\sim$100\,au. Assuming that the three-dimensional
structure of the core resembles a sphere, we can estimate a number
density, and we derive high values for the approximate molecular
hydrogen densities around $6.8\times 10^{10}$\,cm$^{-3}$. The
corresponding column densities are on the order of $10^{25}$\,cm$^{-2}$,
corresponding to visual extinction values around $10^4$\,mag.

\subsection{Emission lines and the dense inner entrained outflow/jet}
\label{emissionlines}

As shown in the spectra in Fig.~\ref{spectra}, we barely find any line
emission signal in these data. On larger scales molecular emission is
expected and also found in previous studies of this region (e.g.,
\citealt{leurini2011,beuther2009c,beuther2017b}), however, this
emission is filtered out by these long baseline observations. In
contrast to this, toward the compact mm peaks the brightness
temperatures of the dust are so high that most lines are found only in
absorption\footnote{At the given high spatial resolution and considering our knowledge about the extended continuum and line emission \citep{beuther2017b}, continuum and line beam filling factors are assumed to be $\sim$1.}. However, there are a few lines seen in Fig.~\ref{spectra}
that are indeed emission lines, even against the strong continuum with
a peak brightness temperature above 1100\,K. This implies that these
spectral line emission features should stem from molecular lines with
very high excitation levels E$_u$/K, most likely above
1000\,K. Identifying these lines is tricky because one finds many
possible line candidates in the typical databases (e.g., JPL
\citealt{pickett1998}, CDMS \citealt{mueller2001}, Splatalogue
https://www.cv.nrao.edu/php/splat/). Table \ref{lines} presents the
frequencies of the emission lines as well as possible transition
candidates. Except for the SiO(5--4) line, we only list line
candidates with excitation temperatures E$_u$/K above 500\,K. We only
list two to three possible candidates for each line, but going through
the line catalogs, others are possible as well. Since we are mainly
interested in the kinematics of the gas in this study, and not so much
in the chemistry, the exact line identification is of lower importance
here, and in the following, we refer to the lines only by their
frequencies.

\begin{table}[htb]
\caption{Emission line parameters}
\label{lines}
\begin{tabular}{lrr}
\hline
\hline
frequency & possible lines & E$_u$/k \\
(GHz)     &               & (K) \\
\hline
217.10 & SiO(5--4) @ 217.10498                        & 31.3 \\
217.82    & HDCO($26_{4,23}-25_{5,20}$) @217.82120           & 1218 \\
          & CH$_3$CH$_2$CN($65_{7,58}-65_{6,59}$) @217.82669 & 980\\
231.896   & C$_2$H$_3$OCHO($96_{11,85}-95_{12,84}$) @231.89506 & 1439 \\
          & CH$_3$CH$_2$CN($57_{8,50}-58_{5,53}$) @231.90159 & 783 \\
232.029   & CH$_2$CHCN($34_{7,28}-35_{6,29}$)v$_{11}$=2 @232.02812 & 1063 \\
          & CH$_2$CHCN($34_{7,27}-35_{6,30}$)v$_{11}$=2 @232.03717 & 1063 \\
232.693   & H$_2$O($5_{5,0}-6_{4,3}$) @232.68670$^a$               & 3462 \\
          & CH$_2$CHCN($28_{2,27}-28_{1,28}$)v$_{15}$=1 @232.68736 & 683 \\
          & (CH$_3$)$_2$CO($43_{16,27}-43_{15,28}$)AA @232.69487 & 686 \\
234.82    & (CH$_3$)$_2$CO($67_{48,20}-67_{47,21}$)EE @234.81810 & 1912 \\
          & HC$_5$N(88--87)v$_{11}$=1 @234.81890 & 655 \\
235.965   & HCOOH($47_{6,42}-46_{7,39}$) @235.96447 & 1337 \\
          & CH$_2$CHCN($63_{3,60}-64_{2,63}$)v$_{11}$=1 @235.96961 & 1281 \\
\hline
\hline
\end{tabular}
~\\
Notes: $^a$ suggested in \citet{ginsburg2018}\\
\end{table}

All compact emission lines can be well imaged and they show mainly
emission around mm1. Since the 217.82\,GHz line is in the spectral
window with the highest spectral resolution (244\,kHz, see Section
\ref{obs}), we do most of the analysis with that line. Other line
detections toward mm2 and mm12 will be discussed in the following
Section \ref{others}.
 
Figure \ref{mom1} presents a first moment map of the 217.82\,GHz line,
and one identifies a clear velocity gradient in approximately
northwest-southeast direction, similar to the general elongation of
the dust continuum emission in this region. For comparison,
Fig.~\ref{channel} shows the corresponding channel map (binned to
2\,km\,s$^{-1}$), and one sees that the highest velocity emission is
peaked at the northwestern and southeastern sides very close to the
continuum peak position whereas the emission in channels around the
$\varv_{\rm lsr}$ is more evenly distributed around the central
continuum peak. In these central channels, the emission is ring-like,
however, that is unlikely to reflect the real molecular gas
distribution but it is probably due to the high optical depth of the
continuum and corresponding self-absorption in that direction. In an
alternative representation of the data, the first moment map in Figure
\ref{mom1} also shows that the highest velocities are close to the
center of mm1 with values of +7.9 and $-9.0$\,km\,s$^{-1}$ within a
separation of only $\sim$0.01$''$. Since a moment map measures
intensity-weighted peak velocities, peak positions of separate
velocity channels can be separated by less than a beam size. A
different measure can be the spatial separation of the peak positions
in the highest-velocity channels shown in Figure~\ref{channel}: The
spatial peak positions in the channels at $-18$ and +14\,km\,s$^{-1}$
are separated by only $0.049''$ or $\sim$108/49\,au at the two
potential distances. Because of the spatially more distributed
emission in the intervening channels, peak positions closer to the
$\varv_{\rm lsr}$ are difficult to measure. Independent of that, the
analysis of the first moment map as well as the channel map shows that
the highest velocities peak close to the continuum emission peak of
mm1.

The general outflow structures in this region have been discussed in
several studies. \citet{leurini2009,leurini2011} suggest three
different outflows whereas \citet{klaassen2015} propose a unified
picture where all features may be explained by one outflow
structure. The recent ALMA CO(6--5) data reveal a red-shifted outflow
emanating from mm1 in northwestern direction \citep{beuther2017b}
which coincides with outflow OF1 from
\citet{leurini2011}. Interestingly, the single outflow suggested by
\citet{klaassen2015} in approximately northeast-southwest direction
has the blue-shifted component toward the north, inverse to what is
found in the CO(6--5) data \citep{beuther2017b}. Therefore, the
interpretation of multiple outflows driven by several sub-sources
appears more likely. This also makes sense in the framework of the
many mm and cm continuum sources identified in this region which are
likely to drive several independent outflows.

\begin{figure}[htb]
  \includegraphics[width=0.49\textwidth]{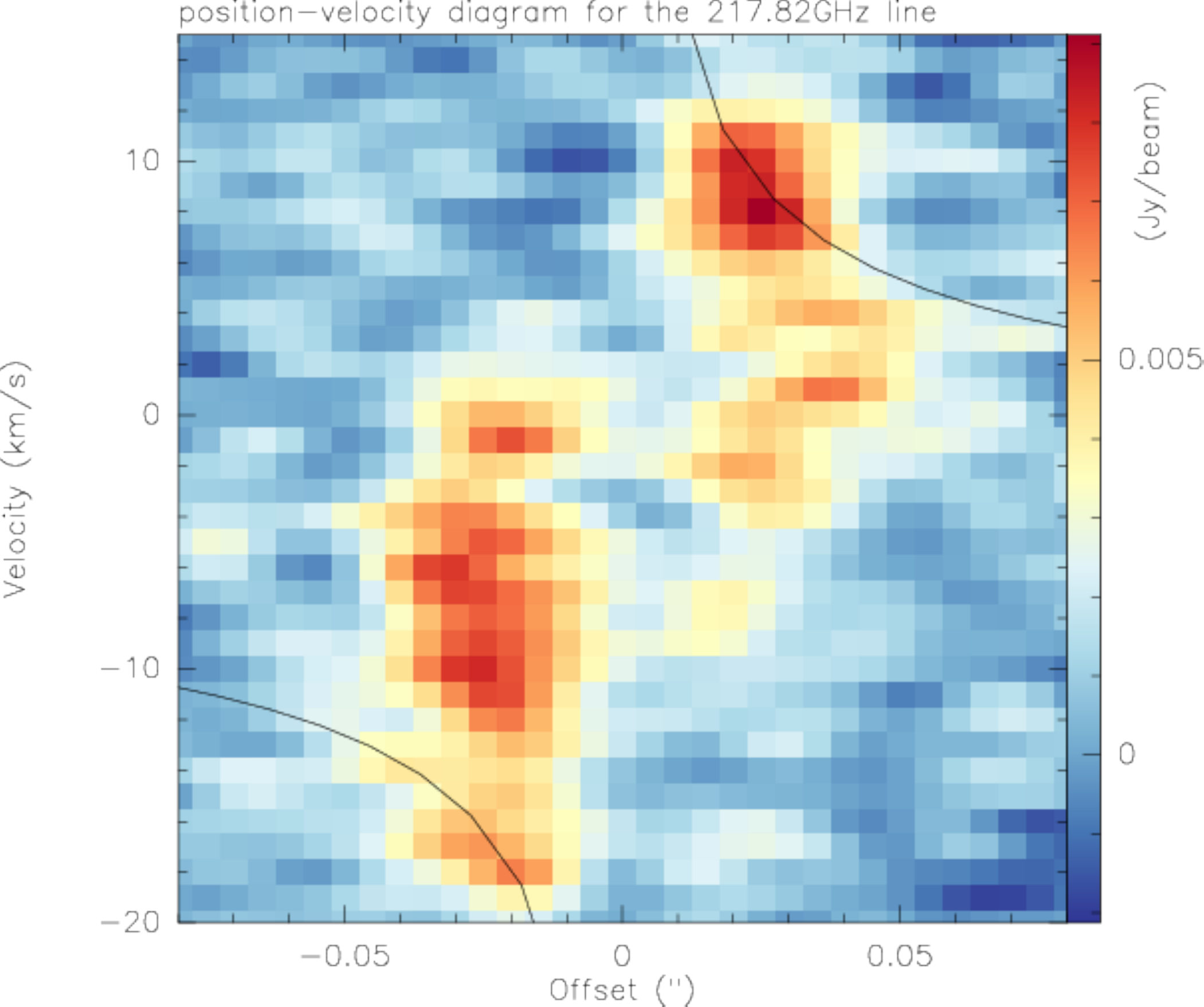}
  \caption{Color-scale shows for mm1 the position-velocity diagram
    for the line at 217.82\,GHz along the axis of the strong velocity
    gradient visible in Fig.~\ref{mom1}. The lines correspond to a
    Keplerian curve around a 10\,M$_{\odot}$ central object, not
    representing the data well.}
\label{pv} 
\end{figure}

The fact that one sees the highest-velocity gas in several spectral
lines close to the center of source mm1 (Keplerian-like signature),
and that the mm continuum emission is elongated in the same direction
as the velocity gradient, suggests that these features may be caused
by a Keplerian disk. However, several observational features argue
against this interpretation. Starting with our knowledge about this
region, as shown in our previous ALMA 690\,GHz data
\citep{beuther2017b}, the molecular outflow originating from mm1 is
directed in the northwest-southeast direction, similar to the velocity
gradient direction in the 217.82\,GHz line. Furthermore, the
rotational motion visible in the CH$_3$CN($37_k-36_k$) lines at
$\sim$680\,GHz with almost Keplerian-like motion is oriented
perpendicular to the outflow and 217.82\,GHz velocity gradient (Fig.~8
in \citealt{beuther2017b}). In addition to this, the position-velocity
diagram of the 217.82\,GHz line along the orientation of the velocity
gradient (Fig.~\ref{pv}) shows the high-velocity gas toward the
center, but one barely sees any lower-velocity gas further out as
would be expected for Keplerian motion.

\begin{figure}[htb]
  \includegraphics[width=0.49\textwidth]{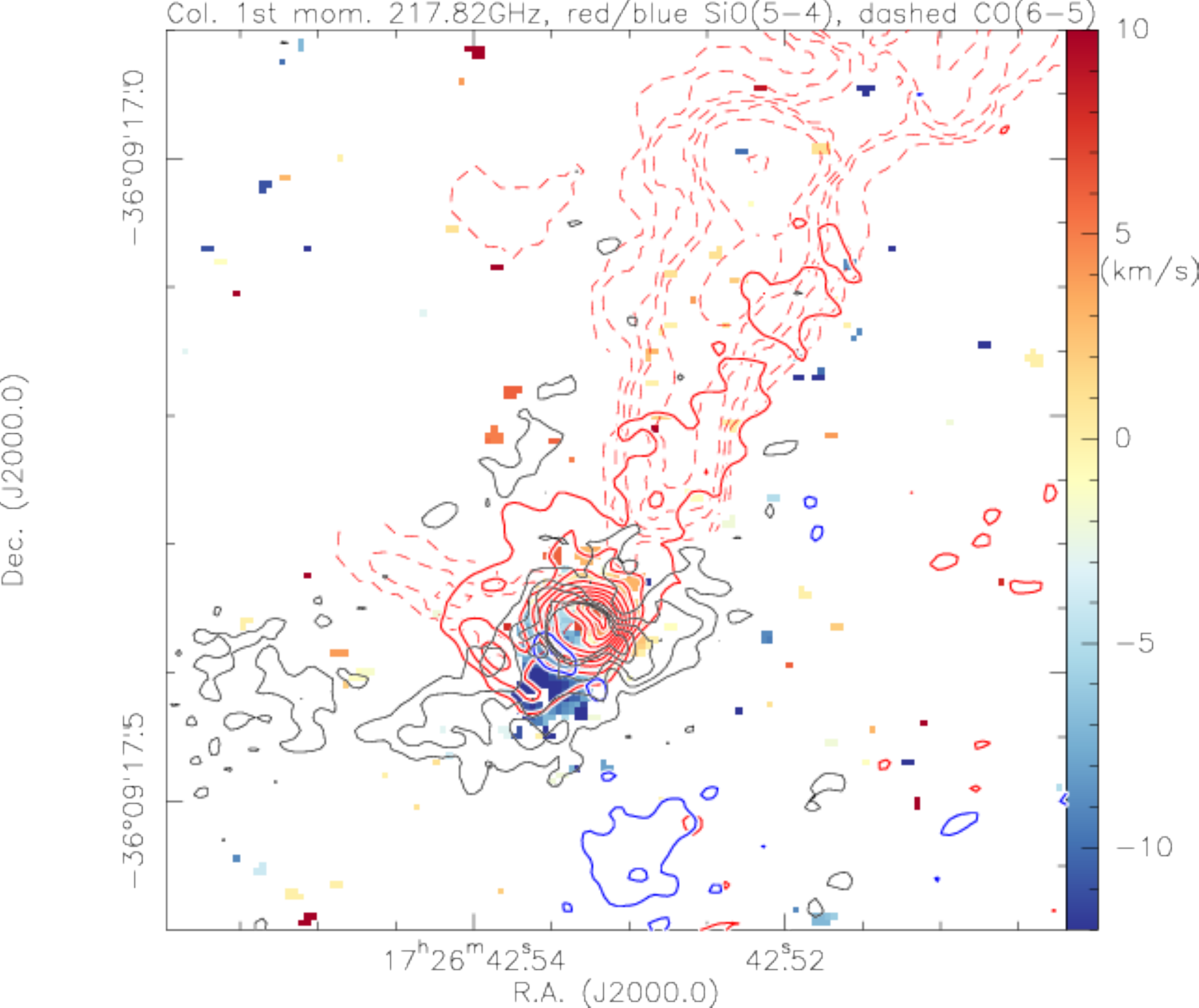}
  \caption{Color-scale shows for mm1 the first moment map
    (intensity-weighted peak velocities) from the line at
    217.82\,GHz. The blue and red solid contours present the blue- and
    redshifted SiO(5-4) emission integrated between [-22,-14] and
    [0,24]\,km\,s$^{-1}$, respectively. The red dashed contours show
    the red-shifted CO(6--5) emission from
    \citet{beuther2017b}. Blue-shifted CO(6--5) can not be properly
    imaged in that dataset, most likely because of missing short
    spacings. The black contours show the lower level 1.3\,mm
    continuum data from the 4$\sigma$ levels of 0.2\,mJy\,beam$^{-1}$
    to 1.0\,mJy\,beam$^{-1}$ in 4$\sigma$ steps.}
\label{mom1_co} 
\end{figure}

For comparison, we checked whether we can see the outflow in our new
SiO(5--4) data as well. Figure \ref{mom1_co} shows the previous
CO(6--5) data \citep{beuther2017b} as well as the blue- and redshifted
SiO(5--4) emission. Although the outflow is typically more extended,
SiO traces a more compact structure that can also be identified in our
data. One sees that the extended redshifted SiO(5--4) emission traces
a collimated jet-like structure emanating from mm1 toward the
northwest that at larger scales the spatially coincides with one
cavity wall that is also seen in the CO(6--5) emission. In addition to
this, the compact blue- and redshifted SiO emission is also oriented
in the northwest-southeast direction around the mm1 source. This
high-velocity SiO gas shows the same spatial and velocity orientation
we also see in the highly excited 217.82\,GHz line.

It is interesting to note that in recent ALMA SiO(5--4) observations
toward the high-mass disk candidate G17.64 at $\sim 0.2''$, the
thermal SiO emission is very compact with a velocity gradient
perpendicular to the outflow \citep{maud2018}. In that case the SiO
emission is interpreted as being emitted from a disk. While thermal
SiO in the past has typically been attributed to shocks from outflows
(e.g., \citealt{schilke1997a}), the G17.64 data indicate also possible
different origins of thermal SiO emission. In that context, we cannot
exclude that some of the compact SiO emission in G351 toward the
center of mm1 may have a disk-contribution. However, in G351
considering also the more extended SiO emission, its spatial and
velocity-association with the CO(6--5) emission, and the perpendicular
velocity gradient found in CH$_3$CN($37_k-36_k$) \citep{beuther2017b},
the SiO emission presented here seems to be dominated by the outflow.

The combination of these various observations indicates that the
highly excited line emission we see in our new long baseline ALMA data
are likely tracing the innermost jet/outflow driven by mm1. It is
interesting to note that the 1.3\,mm continuum emission is also
elongated along the direction of the outflow
(Fig.~\ref{mom1_co}). Therefore, this extended 1.3\,mm emission
apparently does not stem from an inner disk, but seems rather to be
related to the inner jet region of the source as well. Similar
outflow-related dust continuum emission features were found toward the
low-mass outflow region L1157 \citep{gueth2003}. While the
southeastern extension of the dust emission is elongated along the
outflow direction, the northwestern continuum structure more resembles
a cone-like structure often observed in molecular outflows (e.g.,
\citealt{arce2006}). As seen in Fig.~\ref{mom1_co}, these cone-like
continuum features toward the northwest form an envelope around the
central SiO jet-like emission, possibly tracing the outflow cavity
walls.

\begin{figure}[htb]
  \includegraphics[width=0.49\textwidth]{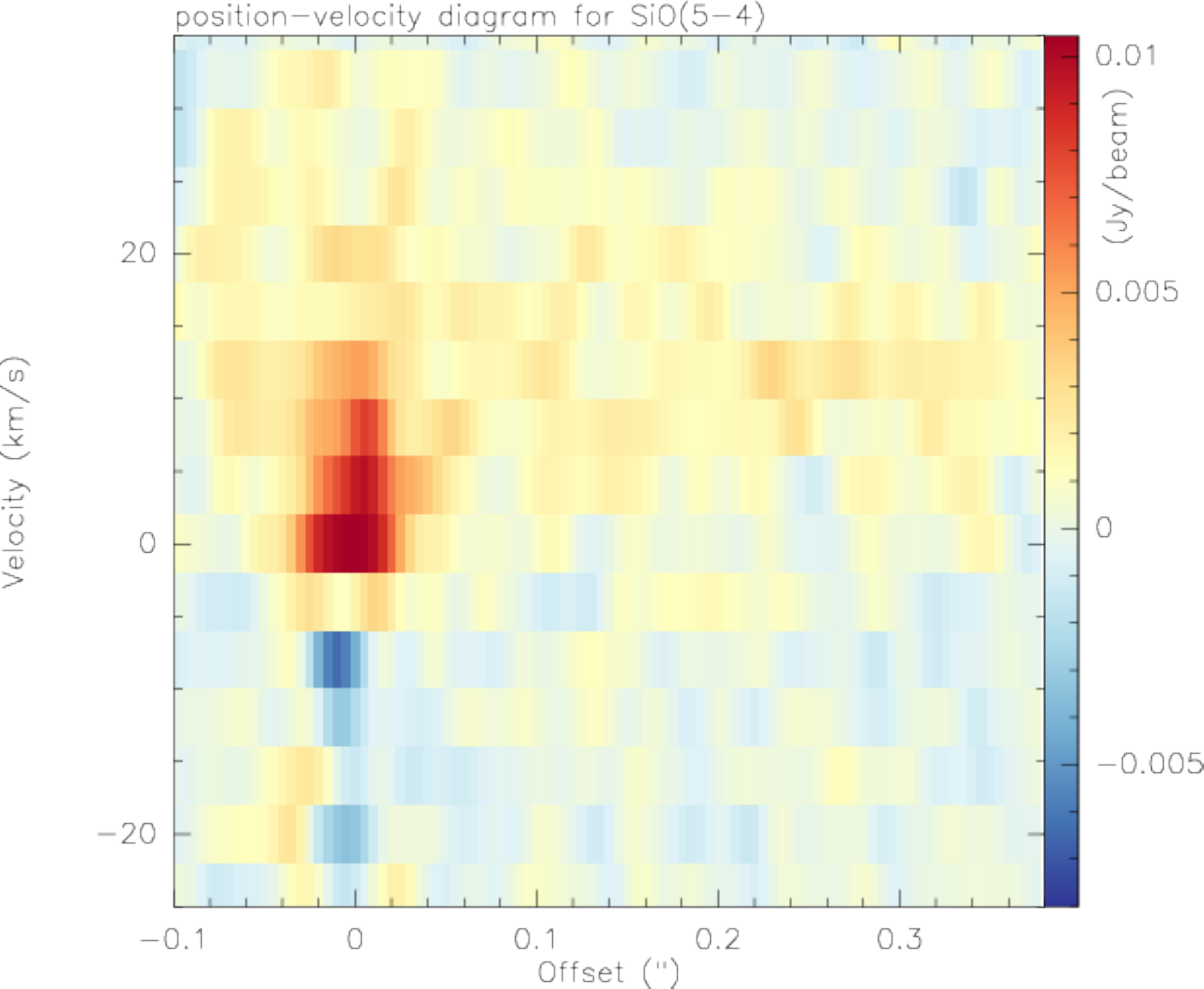}
  \caption{Color-scale shows the position-velocity diagram from
    SiO(5-4) along the outflow axis seen in Fig.~\ref{mom1_co}.}
\label{pv_sio} 
\end{figure}

To compare the larger-scale velocities from SiO(5--4) with the
smaller-scale velocity structure in the 217.82\,GHz line,
Fig.~\ref{pv_sio} shows the position-velocity diagram extracted along
the axis of the outflow. While the blue-shifted SiO emission is
largely filtered out by our extended baseline configuration, we find
SiO(5--4) emission at almost all redshifted velocities close to the
center of mm1. Going further outward in the northwestern direction
(positive offsets in Fig.~\ref{pv_sio}), weaker emission is found
around velocities of $\sim 10$\,km\,s$^{-1}$ but without any strong
gradients. None of these data show a ``Hubble-law'' like velocity
distribution, where the velocities would increase with distance from
the center (e.g., \citealt{downes1999,arce2007}). Rather the opposite,
we find that the highest velocities are close to the
center. Implications for the entrainment process are discussed in
Section \ref{entrainment}.

\subsection{Emission lines in other cores}
\label{others}

\begin{figure*}[htb]
  \includegraphics[width=0.49\textwidth]{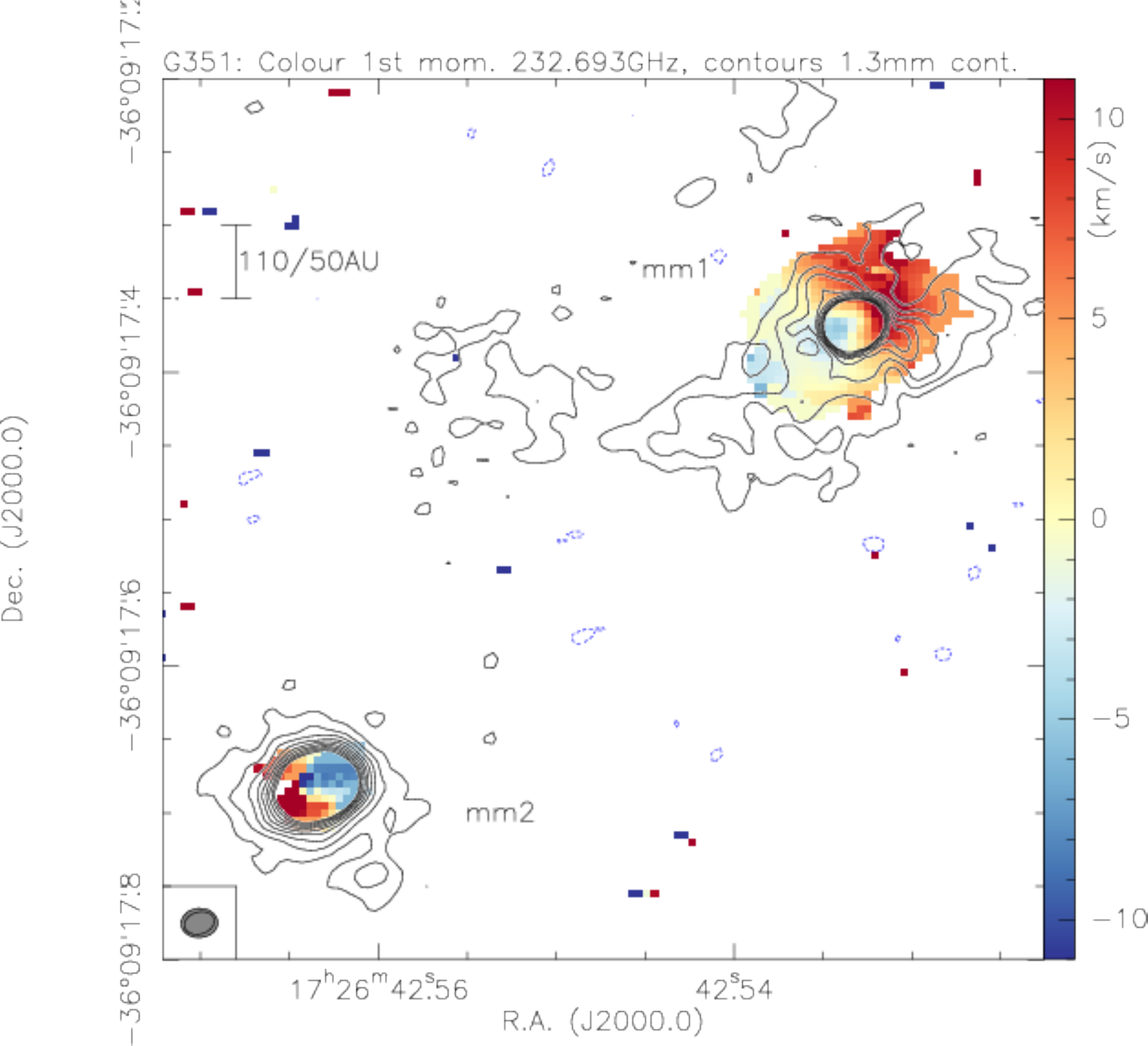}
  \includegraphics[width=0.49\textwidth]{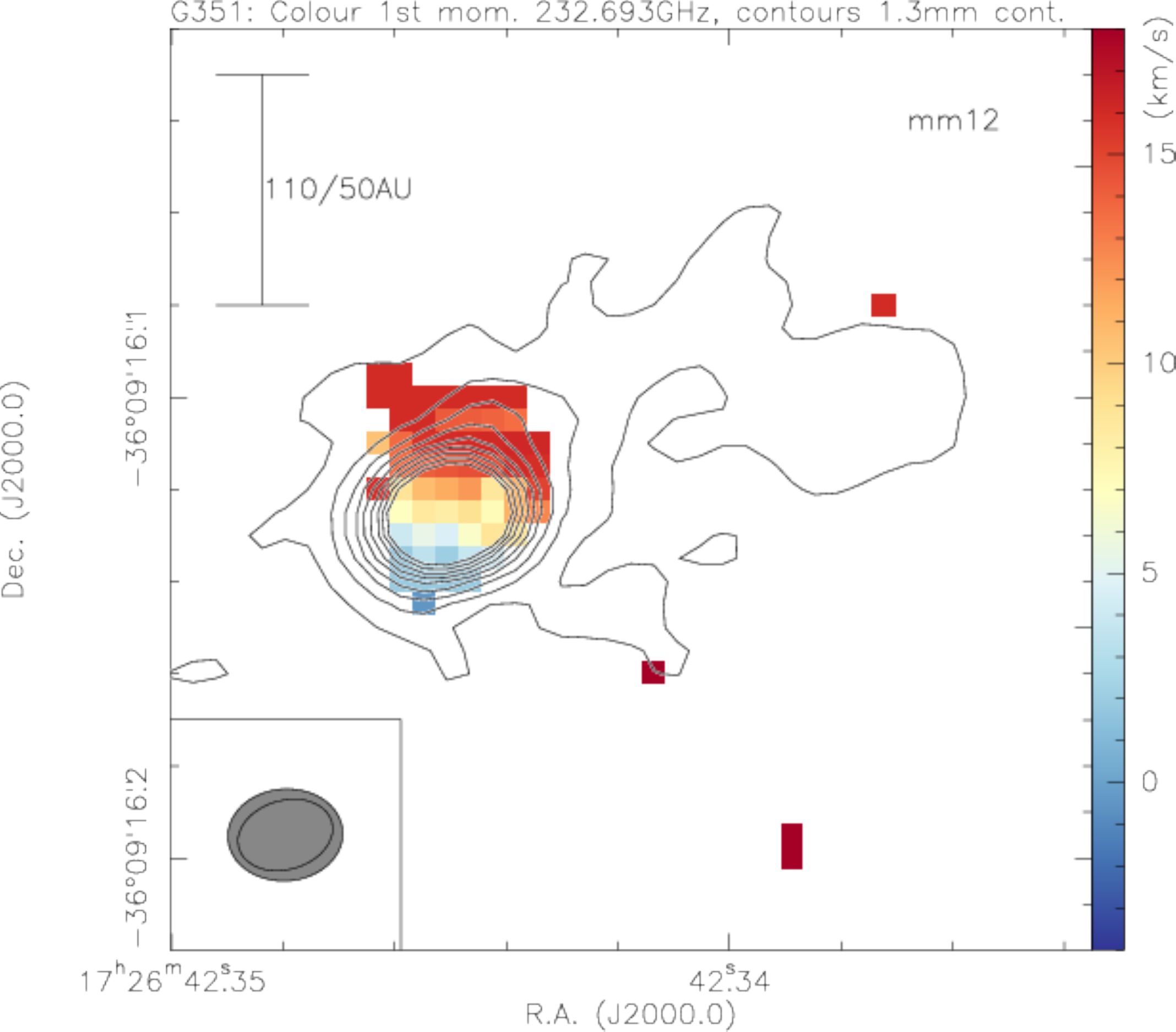}
  \caption{Color-scale shows the first moment maps
    (intensity-weighted peak velocities) from the line at
    232.693\,GHz.  The left panel focus' on mm1 and mm2, whereas the
    right panel presents mm12. The contours are the 1.3\,mm continuum
    data in $4\sigma$ level of 0.2\,mJy\,beam$^{-1}$ up to a
    3\,mJy\,beam$^{-1}$ level (to avoid too many contours in the
    center). A scale-bar and the resolution elements of the line
    (gray) and continuum are shown as well.}
\label{mom1_23269} 
\end{figure*}

For completeness, we also report detections of emission lines toward a
few other sub-sources. We detect line emission from fewer lines toward
mm2 and mm12, while all other sub-sources remain undetected in
emission lines. Figure \ref{mom1_23269} presents the first moment maps
toward mm1, mm2 and mm12 in the emission line at 232.693\,GHz. While
for mm1, the velocity structure previously found in the 217.82\,GHz
line is well recovered, it is interesting that the other two sources
also depict clear velocity gradients. The one in mm2 has approximately
the same orientation as that in mm1, however, the blue- and
red-shifted sides are swapped. For mm12, the velocity gradient is in
the northsouth direction. Since we do not have any information about
molecular outflows from these two sub-sources mm2 and mm12 (the cm
continuum emission from \citet{zapata2008} is not associated with any
of these features either), we cannot infer whether these velocity
gradients are caused by rotation, outflows or any other process.

\begin{figure}[htb]
  \includegraphics[width=0.49\textwidth]{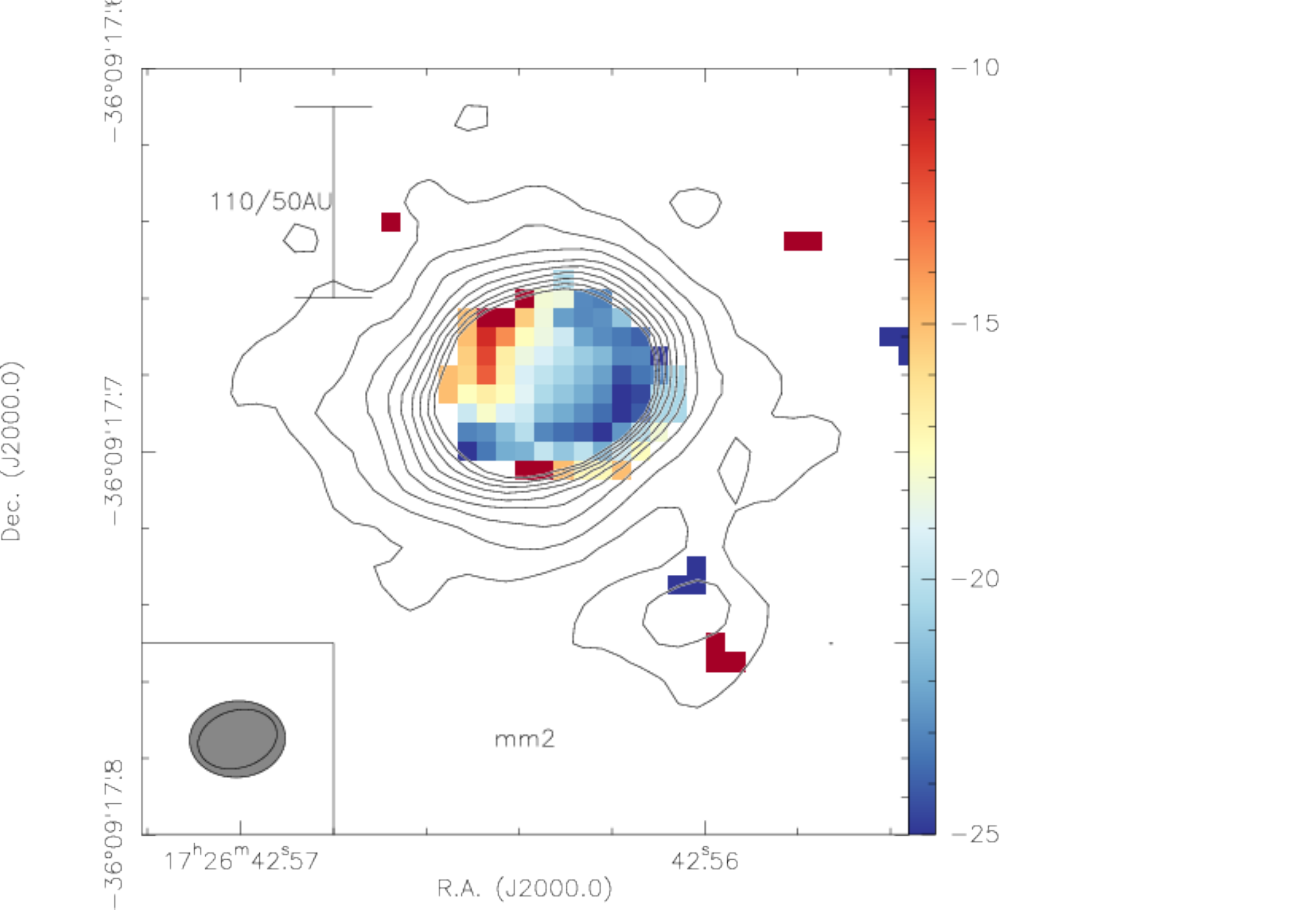}
  \caption{Color-scale shows the first moment map
    (intensity-weighted peak velocities) from the line at 231.896\,GHz
    for the source mm2. The contours are the 1.3\,mm continuum data in
    $4\sigma$ level of 0.2\,mJy\,beam$^{-1}$ up to a
    3\,mJy\,beam$^{-1}$ level (to avoid too many contours in the
    center). A scale-bar and the resolution elements of the line
    (gray) and continuum are shown as well.}
\label{mom1_231896} 
\end{figure}

For mm2, an additional feature arises when imaging the line at
231.896\,GHz (Fig.~\ref{mom1_231896}). While the structure in mm1 is
the same as found in the other lines, toward mm2 we find emission
mainly blue-shifted from the $\varv_{\rm lsr}\sim -3.6$\,km\,s$^{-1}$
to $\sim -20$\,km\,s$^{-1}$. This emission can either be from the same
line, just blue-shifted to high velocities, or it may stem from
another spectral line that we cannot differentiate
properly. Independent of that, the velocity gradient found in this
line is approximately in the northeast-southwest direction, roughly
perpendicular to the velocity gradient seen before in the 232.693\,GHz
line. The perpendicular directions of the velocity gradients in the
two lines toward mm2 indicates that one may trace an outflow/jet
whereas the other may trace rotation. Since we do not have any
additional information at hand, we refrain from further interpretation
of these structures.

\section{Discussion}

\subsection{Fragmentation}
\label{fragmentation}

Fragmentation of molecular clouds and star-forming regions is a
hierarchical process, and one finds fragmentation from large-scale
molecular clouds (e.g.,
\citealt{blitz1993,kramer1998,mckee2007,dobbs2014}) to filament
fragmentation from giant 100\,pc long filaments
\citep{jackson2010,goodman2014,ragan2014,wang2014,wang2015,zucker2015,abreu2016}
to those within star-forming regions (e.g.,
\citealt{andre2010,andre2014,hacar2013,beuther2015b}). It is
interesting to note that there seems to be a difference in the
dominant process driving the fragmentation of filaments and that of
star-forming clumps and cores. While filaments can often be described
by self-gravitating cylinders (e.g.,
\citealt{jackson2010,kainulainen2013,beuther2015b}), star-forming
regions are often well-described by thermal and/or turbulent Jeans
fragmentation (e.g.,
\citealt{palau2013,palau2014,wang2014,beuther2018b}, Svoboda et al.~in
prep.). While previous studies of fragmentation in (high-mass)
star-forming regions typically studied the fragmentation of
star-forming regions from scales of several 10$^4$\,au down to
scales of a few 1000\,au, in this study we now zoom into the
most central high-mass star-forming core and investigate the
fragmentation on scales of hundreds to 1000\,au.

What would be the thermal Jeans fragmentation properties of a typical
high-mass core of several 1000\,au size? Using the study of the IRAM
large program CORE of 20 high-mass star-forming regions,
\citet{beuther2018b} find, for regions below 2\,kpc distance, typical
mean volume number densities between $10^7$ and $10^8$\,cm$^{-3}$
averaged over spatial resolution elements of $\sim
0.3''-0.4''$. Assuming mean temperatures of around 50\,K for these
regions because the star formation process is already active, we find
Jeans lengths between roughly 870 and 2800\,au, and Jeans masses
between 0.1 and 0.4\,M$_{\odot}$. Although our observed separations
and masses are uncertain due to the poorly constrained distance,
projection effects as well as missing flux and optical depth
uncertainties, the typical observed source separations and masses
(Section \ref{sec_cont}) are on the order of these Jeans estimates.

One may also ask whether all the identified twelve sub-sources will
form stars. While the compact structures like mm1, mm2, mm4 or mm12
seem good candidates for actual forming protostars (mm1 is actively
driving an outflow, mm2 and mm12 are also showing line emission from
highly excited lines), other structures like mm5, mm7 or mm11 appear
more diffuse and may simply be local over-densities which could be
transient structures and will not form stars within them. In fact,
these more diffuse structures may be parts of the otherwise largely
filtered out envelope which, if it were infalling, could result in
these structures ending up on some protostar later in its evolution.

In this context, the spatial structure of mm9 appears intriguing. Its
partly bent and wave-like structure has an appearance similar to
gravitationally interacting core structures also observed in recent 3D
radiation-hydrodynamical simulations of high-mass star-forming regions
(e.g., \citealt{meyer2018}; Ahmadi et al.~in prep.). For example,
\citet{meyer2018} model an infalling envelope around a
self-gravitating disk that evolves while being irradiated by the
time-dependent evolution of the central protostar. Independent of
their initial angular velocity distribution, all their models form
similarly bent-like structures at scales of several hundred to
thousand au from the center.  From the observational side, infalling
streamer-like structures have recently also been reported in other
high-mass star-forming regions (e.g.,
\citealt{maud2017,izquierdo2018,goddi2018}). In particular, the
observations and analysis of the high-mass region W33A reveals
comparably extended structures at $\sim$1000\,au separation from the
center that are attributed to spiral-like infalling gas streamer
\citep{maud2017,izquierdo2018}. If those data were observed with
poorer uv-coverage, these more extended spiral-like structure may well
resemble what we see here in G351.77-0.54. Hence, within the
G351.77-0.54 region, so far, we do not find a larger disk-like
structure like for example in AFGL4176 \citep{johnston2015} or G17.64
\citep{maud2018}, but it is more a multisource environment with
smaller-scale structures like those found in W33A or W51n (e.g.,
\citealt{izquierdo2018,goddi2018}).

However, with the given data, such interpretation is for G351.77-0.54
still largely premature and speculative. To investigate this
``inter-core'' gas in more detail, complementary ALMA observations
with shorter baselines are needed. With such data, not only can the
more diffuse dust emission be studied, but also the line emission at
lower brightness to investigate the kinematics of the connecting gas
structures.

\subsection{Turbulent entrainment}
\label{entrainment}

Following the interpretation outlined in Section \ref{emissionlines}
that the compact high-energy line emission stems from the innermost
jet/outflow driven from mm1, we can use these data to further analyze
the entrainment mechanism.  Since large-scale molecular outflows are
typically entrained gas and do not directly trace the underlying
driving jet, gas entrainment processes are important in the
star-formation process. Jet-outflow entrainment processes have been
discussed for many years (e.g.,
\citealt{masson1993,raga1993,stahler1994,downes1999,arce2002,canto2003,arce2007,frank2014,raga2016}). One
of the key-diagnostics to study gas entrainment are position-velocity
diagrams, and one recurring feature in the literature is the so-called
Hubble-law of molecular outflows where the observed gas velocities
increase with increasing distance from the source (e.g.,
\citealt{raga1993,downes1999,arce2007}). The most discussed outflow
entrainment scenarios are turbulent entrainment, jet-bow-shock
entrainment as well as wide-angle winds (e.g.,
\citealt{canto1991,masson1993,stahler1994,chernin1995,bence1996,cliffe1996,li1996,hatchell1999,lee2001,arce2002,arce2007,frank2014}). A
summary of these processes and the corresponding diagnostics, in
particular the position-velocity diagrams for the various processes
can be found in \citet{arce2007}. While the prompt-entrainment
jet-bow-shock models and the wide-angle winds both predict velocity
structures where the velocity increases with distance to the source,
the more steady-state turbulent entrainment predicts rather the
opposite with high velocities close to the source (Fig.~2 in
\citealt{arce2007}). Interestingly, many outflow studies on larger
spatial scales seem to prefer the jet-bow-shock and wide-angle models
(e.g.,
\citealt{yu1999,lee2002,arce2002,arce2007,ray2007,tafalla2017}).

However, although the central peak position around mm1 suffers from
absorption (e.g., Figs.~\ref{spectra} and \ref{channel}), our new data
of the hot, likely entrained gas around mm1 do not show any
Hubble-like velocity structure with increasing velocity away from the
center, but we clearly find high-velocity gas very close to the center
of the source (Section \ref{emissionlines}, and Figs.~\ref{mom1} and
\ref{channel}). Furthermore, also the position-velocity diagram of
SiO(5-4) along the outflow axis exhibits the highest velocities close
to the center and then almost constant velocities with increasing
distance from the center (Fig.~\ref{pv_sio}). Hence, also out to
distances of several hundred au from the center of mm1, the new ALMA
long baseline data do not show Hubble-like velocity-structure. We
stress that to our knowledge barely any high-mass jet study in thermal
line emission so far has ever resolved the scales below 50\,au from
the driving protostar as we do here (a notable exception is source I
in Orion, \citealt{ginsburg2018}). Hence, we reach new territory with
the supreme capabilities of ALMA. Compared to the basic predictions of
the prompt bow-shock entrainment models at the heads of the jets, and
the steady-state turbulent entrainment models via Kelvin-Helmholtz
instabilities along the jets (e.g.,
\citealt{bence1996,arce2002,arce2007}), on these small scales, the
position-velocity diagnostics are more consistent with turbulent
entrainment at the jet-outflow interface. From a physical point
  of view, both processes have to exist, steady-state turbulent
  entrainment via Kelvin Helmholtz instabilities along the
  jet-envelope interface as well as prompt entrainment at the head of
  jets via bow-shocks. Therefore, the question is not that much about
  the existence of the different processes but rather which of them
  dominates in one or the other source and at which spatial
  scales. Hence, the results for G351.77-0.54 do not necessarily
contradict previous observational results that found the Hubble-like
velocity structure on larger spatial scales. It is rather
possible that on the smallest scales, as observed here, turbulent
entrainment may be more important whereas on larger scales bow-shock
entrainment could become the dominant process. This issue can
again only be directly tackled by observations that connect all
spatial scales, from those observed in this study of G351.77 to the
larger scales of the more typically observed outflows.

An additionally interesting feature with respect to the inner
jet/outflow and accretion disk is the extension of the 1.3\,mm
continuum emission along the axis of the jet/outflow. While
elongations of 1.3\,mm continuum emission are typically attributed to
the dense cores and potential accretion disks, this seems less likely
for the case of G351.77-0.54 here. As outlined in Section
\ref{emissionlines} and Fig.~\ref{mom1}, while toward the southeast of
mm1 the 1.3\,mm continuum emission is elongated along the outflow
axis, toward the northwest, the mm emission rather resembles a
cone-like structure almost enveloping the central SiO(5--4) jet and
CO(6--5) outflow emission. While we do not have a reasonable
explanation for this asymmetry of the 1.3\,mm emission, both features
appear qualitatively consistent with entrainment of the ambient dust
and gas. It may be that for the blue wing, we see into the outflow
lobe and trace the more compact structures, whereas the red wing could
cover the central jet better and one would see more of the
limb-brightened edges. As discussed in Section \ref{fragmentation},
typical mean densities in these dense cores are $10^7$ to
$10^8$\,cm$^{-3}$ (e.g., \citealt{beuther2018b}), and close to the
peak positions even higher (Section \ref{sec_cont}). At such high
densities, gas and dust are well coupled, and if the gas is entrained
by the underlying jet, it should be no surprise that similar spatial
structures are also observed in the dust continuum emission.

\section{Conclusions}
\label{conclusion}

Resolving for one of the first times a high-mass star-forming region
at sub-50\,au spatial scales (18 or 40\,au resolution depending on the
distance of 1.0 or 2.2\,kpc) reveals various highly interesting
outcomes. In the dust continuum emission, we resolve twelve distinct
sub-sources within a region of roughly $6''\times 6''$ or
6000/13200\,au at 1.0/2.2\,kpc distance. Projected separations between
sub-sources range from more than 1000\,au to several 100\,au but can
be even smaller below 100\,au. Since we filter out the large-scale
emission, very extended structures are not visible in the data, but
the largest features we can identify are between approximately 200 and
600\,au in size, similar to the size-scales predicted by theoretical
simulations for disks around embedded high-mass protostars. Brightness
temperatures toward the main continuum peak positions are in excess of
1000\,K, indicating high optical depth toward these positions. This
high optical depth as well as the missing flux only allows us to
derive lower limits for the masses and column densities of the
sub-sources. In the hierarchical picture of fragmentation from
large-scale clouds to small-scale cores, we find that the
fragmentation properties of this high-mass core with a size of several
1000\,au is consistent with thermal Jeans fragmentation.

The high continuum optical depth is also confirmed by the fact that
most spectral lines toward the dust peak positions are only seen in
absorption. However, we find a few lines in emission toward the main
source mm1 as well as two other sources mm2 and mm12. Because of the
high dust continuum brightness temperatures, these emission lines have
to stem from transitions with very high excitation levels $E_u/k$ on
the order of 1000\,K. The emission toward mm1 shows a velocity
gradient with the highest velocities found toward the core
center. While this high-excitation line emission by itself may be
indicative of a Keplerian disk-like structure, additional information
let us favor a different interpretation. In particular, the elongation
of this velocity gradient is exactly along the direction of the
molecular outflow observed in SiO(5--4) and CO(6--5) and perpendicular
to the rotational velocity gradient previously identified in 690\,GHz
CH$_3$CN$(37_k-36_k)$ data \citep{beuther2017b}. Hence this gas
emission does likely not arise from an inner accretion disk but is
more consistent with being caused by the central jet/outflow
region. The high-velocity gas close to the center is consistent with
steady-state turbulent entrainment of the molecular gas from the
underlying jet via Kelvin-Helmholtz instabilities. So far, in
G351.77-0.54 we do not identify a larger-scale disk but the region is
composed of a smaller-scale multisource environment.

Since this study only covers the smallest spatial scales without any
baselines below 55\,m, we are not seeing any extended emission. Such
extended emission would be important to connect the scales from
thousands and more au down to sub-50\,au scales observed here. How
does the dense compact structure connect to the more extended gas? How
are the kinematic connections from large to small spatial scales?
These questions can only be addressed by data with a more complete
uv-coverage that would trace all spatial scales and would also be
sensitive to lower brightness emission from dust and gas. These
investigations will be conducted in future studies.

\begin{acknowledgements} 
  The proposal and design of this study was devised in close
  collaboration with Malcolm Walmsley who unfortunately passed away in
  2017. We like to express our deep gratitude to Malcolm for many
  years of close collaboration, which was always a delight and very
  inspiring! We also like to thank a lot an anonymous referee for
  detailed comments improving the paper. This paper makes use of the
  following ALMA data: ADS/JAO.ALMA\#2015.1.00496.S. ALMA is a
  partnership of ESO (representing its member states), NSF (USA) and
  NINS (Japan), together with NRC (Canada) and NSC and ASIAA (Taiwan)
  and KASI (Republic of Korea), in cooperation with the Republic of
  Chile. The Joint ALMA Observatory is operated by ESO, AUI/NRAO and
  NAOJ. H.B.~acknowledges support from the European Research Council
  under the Horizon 2020 Framework Program via the ERC Consolidator
  Grant CSF-648505. R.K.~acknowledges financial support from the Emmy
  Noether Research Program, funded by the German Research Foundation
  (DFG) under grant number KU 2849/3-1.
\end{acknowledgements}


\end{document}